\documentclass[letterpaper,12pt]{article}
\usepackage[centering, top=1.0in, bottom=1.0in, left=1.0in, right=1.0in]{geometry}
\usepackage{amsmath}
\usepackage{graphicx}
\usepackage[super]{nth}
\usepackage{natbib}
\usepackage[capposition=top]{floatrow}
\usepackage[T1]{fontenc}
\usepackage{mdwlist}
\usepackage{amsfonts}
\usepackage{tabularx}
\usepackage{amssymb}
\usepackage{amsthm}
\usepackage{lscape}
\usepackage[bf]{caption}
\usepackage{subcaption}
\usepackage{verbatim}
\usepackage[usenames,dvipsnames]{xcolor}
\usepackage{setspace}
\usepackage{arydshln}
\usepackage{array}
\usepackage{pdfsync}
\usepackage[hang,flushmargin]{footmisc} 
\usepackage{afterpage}
\usepackage{booktabs,dcolumn,tipa}
\usepackage{epigraph}
\usepackage{hyperref}
\hypersetup{colorlinks=true, linkcolor=NavyBlue, citecolor=NavyBlue, urlcolor=NavyBlue}
\usepackage{float}
\usepackage[small,compact]{titlesec}
\titleformat{\chapter}
 {\normalfont\Large\bfseries}{\thechapter}{1em}{}
\titlespacing*{\chapter}{0pt}{1.5ex plus 1ex minus .2ex}{1.3ex plus .2ex}

\usepackage{subcaption}
\usepackage{longtable}
\usepackage[nottoc,numbib]{tocbibind}
\setcitestyle{aysep={}}
\usepackage[titletoc,title, page]{appendix}

\usepackage[T2A,T1]{fontenc}
\usepackage[utf8]{inputenc}
\usepackage{amsmath,amssymb}
\usepackage[russian,english]{babel}
\usepackage{graphicx}
\usepackage{tabularx}
\usepackage{xcolor}

\begin{document}

\title{\textbf{Artificial Intelligence, Surveillance, and Big Data}\footnote{This working paper is expected to appear in: Diginomics Research Perspectives: The Role of Digitalization in Business and Society (Springer Nature, editor: Lars Hornuf).}}
\bigskip

\author{
\textbf{\href{https://orcid.org/0000-0002-4327-4269}{\includegraphics[scale=0.06]{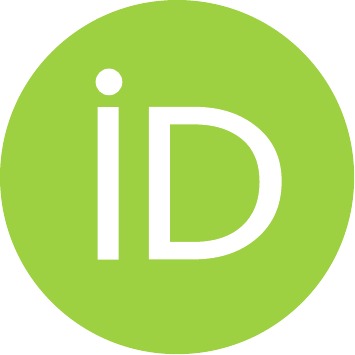}\hspace{1mm}}David Karpa}\thanks{dkarpa@uni-bremen.de}\\
\normalsize University of Bremen  \\
\and
\textbf{\href{https://orcid.org/0000-0002-3059-1150}{\includegraphics[scale=0.06]{orcid.pdf}\hspace{1mm}}Torben Klarl}\thanks{tklarl@uni-bremen.de}\\
\normalsize University of Bremen / \\
\normalsize SPEA, Indiana University, Bloomington
\and
\textbf{\href{https://orcid.org/0000-0001-8652-8874}{\includegraphics[scale=0.06]{orcid.pdf}\hspace{1mm}}Michael Rochlitz}\thanks{michael.rochlitz@uni-bremen.de}\\
\normalsize University of Bremen /\\
\normalsize ICSID, Higher School of Economics, Moscow\\}

\date{\today}

\maketitle

\begin{abstract}
The most important resource to improve technologies in the field of artificial intelligence is data.
Two types of policies are crucial in this respect: privacy and data-sharing regulations, and the use of surveillance technologies for policing.
Both types of policies vary substantially across countries and political regimes.
In this paper, we examine how authoritarian and democratic political institutions can influence the quality of research in artificial intelligence, and the availability of large-scale datasets to improve and train deep learning algorithms.
We focus mainly on the Chinese case, and find that -- ceteris paribus -- authoritarian political institutions continue to have a negative effect on innovation.
They can, however, have a positive effect on research in deep learning, via the availability of large-scale datasets that have been obtained through government surveillance.
We propose a research agenda to study which of the two effects might dominate in a race for leadership in artificial intelligence between countries with different political institutions, such as the United States and China. 
\noindent 

\end{abstract}

\bigskip
%\vspace{15mm}
\vfill
\noindent \textbf{Keywords:} artificial intelligence, political institutions, big data, surveillance, innovation, China

\bigskip

\thispagestyle{empty}

\newpage
\pagenumbering{arabic}
\setcounter{page}{1} % start page number counting over on first page of text

\renewcommand{\baselinestretch}{1.0}
\large \normalsize \setlength{\footnotesep}{\ht\strutbox}

\section{Introduction}\label{intro}

During the late 2000s and early 2010s, a number of publications on recursive learning in multi-layered neuronal networks led to a breakthrough in the field of artificial intelligence \citep{Hinton.2006, Hinton_Osindero_Teh_2006, Krizhevsky_2009, Krizhevsky_Sutskever_Hinton_2012}.
The studies laid the foundation for a new approach to machine learning: recursively self-improving algorithms are structured in layers to create programmable neural networks, which are then used to sort through large amounts of data in a process called ``deep learning''.
Other than with traditional machine learning, deep learning algorithms do no longer require outside training. 
Instead, they are able to autonomously reach a goal set at the beginning, with the speed of learning depending on the amount of raw data available. 

After a couple of years, machine-learning techniques based on deep learning have led to a number of major breakthroughs in various fields such as image, speech and facial recognition, medicine, particle physics, neuroscience and language translation \citep{LeCun_Bengio_Hinton_2015}.
The larger public became aware of the new technology when a team from Deep Mind, an AI firm now acquired by Google, managed to beat Lee Sedol, one of the highest ranking professional Go players, in a five-game match in March 2016 \citep{Silver_et_al_2016}.
While for the game of chess IMB's Deep Blue had already managed to beat world champion Garry Kasparov in 1997, Go is an order of magnitudes more difficult to solve than chess.\footnote{While Deep Blue beat Garry Kasparov in 1997 by using brute force computing, this approach does not work for Go, a game with 250 possible moves each turn, and a typical game depth of 150 moves, resulting in about \(250^{150}\), or \(10^{360}\) possible moves -- more than the number of atoms in the observable universe.}  
Winning the game against one of the world's leading players thus demonstrated the potential of the new technology \citep{Silver_et_al_2017, Bory_2019}. 

Deep Mind's win against Lee Sedol had also another important implication. 
Go was invented in China, is mostly played in China, Japan and Korea, and is still considered by many people in East Asia as a quintessential Asian game \citep{Moskowitz_2013}.
That a Western company managed to win the game with the help of a computer program against a leading Asian player led to a spike in interest in artificial intelligence research in China, triggering a large-scale government program to promote and support the already fast-developing sector.\footnote{\citet[page 3]{Lee.2018} compares the effect of Deep Mind's win against Lee Sedol to America's Sputnik moment: ``Overnight, China plunged into an artificial intelligence fever. The buzz didn't quite rival America's reaction to Sputnik, but it lit a fire under the Chinese technology community that has been burning ever since''.}
5 years later, China has become -- together with the United States -- one of the two leading nations in the field of AI research, with both countries engaged in a tight race for leadership.

\begin{figure}[htbp]
\centerline{\includegraphics[scale=0.5]{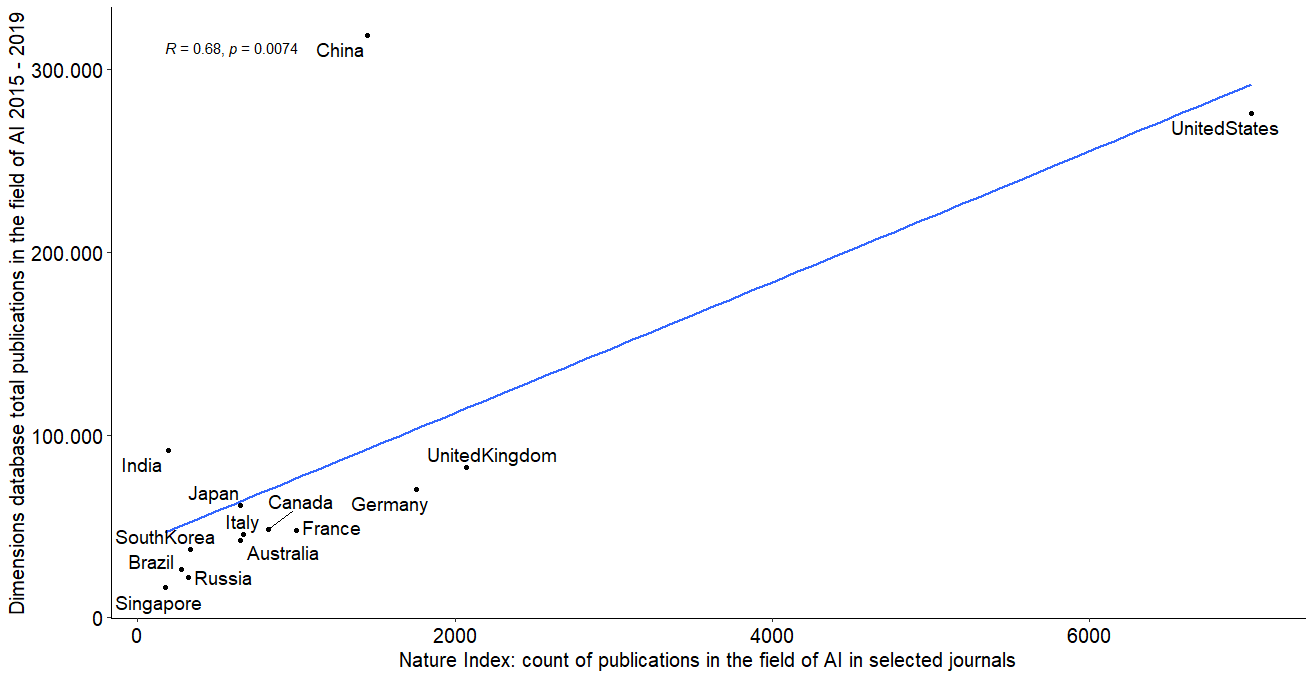}}
\caption{Publications in selected journals (x-axis); publications total (y-axis)} 
\label{pubs}
\end{figure}

Figure \ref{pubs} shows that while China has already overtaken the United States in terms of the overall quantity of AI-related research publications, it still lags behind with respect to the number of high quality publications in applied AI research, where China today is at about the same level as Germany or the UK.\footnote{As a proxy for research quality in artificial intelligence, we use the Nature Index for the year 2020. The index counts publications in applied artificial intelligence published in Nature and its sub-journals, such as Science, the Proceedings of the National Academy of Sciences (PNAS), and others. The list can be accessed under \url{https://www.natureindex.com/faq}, and the source of our data can be found here: \url{https://www.natureindex.com/supplements/nature-index-2020-ai/tables/countries}. As a proxy for the overall quantity of publications in artificial intelligence, we use the Nature Index Dimensions database, which counts all publications in AI from a specific country between 2015 and 2019: \url{https://www.natureindex.com/supplements/nature-index-2020-ai/tables/dimensions-countries}.} 
A similar picture -- higher quantity but still lower quality, on average -- also exists with respect to patents, for example in the case of the internet of things \citep{Kshetri.2017}, and when comparing research institutions.
Here American and European universities are still ahead when looking at high-quality output, whereas Chinese institutions have recently overtaken all other countries in the world with respect to the overall number of AI-related publications per university (\cite{Savage.2020}, see also tables \ref{nature_index_table} and \ref{dimensions_data_table} in section \ref{positive_effects}).

In this paper, we try to understand if China has the potential to overtake the United States also in terms of research quality. 
To answer this question, we focus on one fundamental difference between both countries -- the difference between authoritarian and democratic institutions. 
While the United States are a competitive democracy, China is an authoritarian single-party state.
Our objective is to study how this difference in political institutions can affect the quality of research and innovation in the field of artificial intelligence.
Apart from permitting us to better understand the AI race between the United States and China, this will also allow us to gain a better understanding of the role political institutions play with respect to research and innovation. 

Conventionally, authoritarian political institutions are seen as having a mostly negative effect on innovation (see for example \citealt{Huang_1999, Stokes_2000, Josephson_2005, Tebaldi_2008, Knutsen_2015, Silve_2018, TangTang_2018, Rosenberg_2020}).
In authoritarian political systems, the story goes, censorship, surveillance and repression, while necessary for the survival of the regime, limit the free flow of information, and make it -- on average -- more difficult for scientists to freely engage in research, thus limiting their creativity and inventiveness.\footnote{A certain amount of censorship, surveillance and government control can of course also be found in democratic political systems. We therefore assume a continuum of repressiveness, from complete freedom to complete state control, with the negative effects on creativity increasing with growing government control.}

The central argument of our paper is that with respect to innovation in the field of artificial intelligence, this conventional wisdom might no longer hold, with the negative effects of authoritarian institutions potentially being offset by the opportunities offered through big data.
As pointed out above, the most important input for deep learning algorithms to self-improve are large amounts of data \citep{Hey_2009, Halevy2009, Domingos_2015, Sun.2017, Beraja_Yang_Yuchtman_2021}.
With respect to the availability of large data sets, modern authoritarian states with sophisticated surveillance systems might have a significant advantage over democratic systems, where -- on average -- policies on government surveillance and data sharing are stricter than in autocracies \citep{Kshetri.2014, Chen.2017, Strittmatter_2020, Zeng_2020}.
Less strict data sharing laws can permit authoritarian governments to allocate data gathered through surveillance as an input-subsidy to domestic IT firms, an industrial policy the Chinese government is already increasingly pursuing \citep{Lee.2018, Beraja_Yang_Yuchtman_2021}. 
A side-effect is that in this way, authoritarian states can use the data they have gathered through surveillance to further improve their AI-based surveillance capabilities, thus initiating a positive feedback loop. 
As a result of this process, China has already become a leader in the field of surveillance technologies, and has started to export them to other authoritarian countries \citep{Feldstein_2019}.\footnote{See also ``The Panopticon is Already Here'', The Atlantic, September 2020, \url{https://www.theatlantic.com/magazine/archive/2020/09/china-ai-surveillance/614197/}.}

Our hypothesis is that whether China will manage to catch up to, and potentially overtake the United States to become world leader with respect to cutting-edge innovation in AI will be determined by the trade-off between the negative effects of censorship, surveillance and control on innovation and creativity, and the positive effects of having access to better and more data than countries that do not rely to the same extent on digital surveillance.
As we argue in section \ref{negative_effects}, for the time being -- and for the foreseeable future -- China will remain at a disadvantage when it comes to conventional innovation, as its authoritarian institutions are hindering the free flow of information, as well as intrinsic motivation, creativity, and the possibility for scientists to freely determine their research agenda.  
On the other hand, China has already today an advantage with respect to using big data as an input in AI research, as we document in section \ref{positive_effects}.
We argue that who will win the race in artificial intelligence will ultimately depend on what type of research will be more important in the field during the next couple of years. 
If new and innovative ideas lead to another breakthrough after the deep learning revolution, the US might keep the lead in the foreseeable future. 
If however large datasets to train and improve existing algorithms turn out to be most important during the next decade, China might well become the overall leader in the field. 

This paper is structured as follows. 
Section \ref{theory} introduces a simple theoretical framework, which will be used in the subsequent sections to guide our argument.
The framework consists of three \textit{actors} -- the authoritarian regime, the public, and high-tech firms, a number of \textit{strategies} used by the authoritarian regime to ensure its political survival (repression, censorship, surveillance and public spending), and a number of \textit{outcomes} we are interested in, most importantly the amount of data available to research institutions, and the quality of research and innovation in AI.
Section \ref{negative_effects} discusses the negative effects of censorship and surveillance on research and creativity, thus reiterating and illustrating the conventional argument that on average, authoritarian political institutions have a negative effect on innovation. 
Section \ref{positive_effects} then introduces our argument, by showing how large amounts of data gained through government surveillance can, potentially, have a positive effect on innovation in AI. 
Finally, section \ref{conclusion} summarizes our argument and concludes.

\section{Theoretical framework}\label{theory}

An extensive literature has examined the strategies used by authoritarian governments to remain in power, with \citet{Egorov.2020} providing a recent overview.
According to \citet{Svolik_2009}, authoritarian regimes during the last 70 years have faced two principal challenges -- revolution from below, and palace coups from within. 
Traditionally, most authoritarian regimes have reacted with either repression or cooptation to these challenges.
While repression to get rid of regime opponents has taken the form of arrests, purges, deportation or worse \citep{Moore_1998, Gregory_Schroeder_Sonin_2011, Blaydes_2018, Montagnes_Wolton_2019, Buckley_et_al_2021}, public spending has been used to either convince the general public that it is better off under the status quo than under an alternative regime \citep{Acemoglu.2006}, or to co-opt a more narrowly defined elite, in order to prevent a coup \citep{Bueno_de_Mesquita_2003, Gallagher_Hanson_2015}. 

More recently, the growing importance of the internet and more sophisticated censorship, surveillance and propaganda techniques have led to the emergence of a new type of dictatorship. 
In these so-called ``informational autocracies'', the manipulation and targeted use of information has become the preferred strategy of the government, while repression is no longer as important as before \citep{Guriev_2019, Guriev_2020}.
Examples of modern informational autocracies include such diverse countries as Russia under Vladimir Putin, Hungary under Victor Orban, Turkey under Recep Tayyip Erdoğan, as well as contemporary China.\footnote{As China does not have elections and is still using repression, it is sometimes not seen as a classic ``informational autocracy'' in the sense of \citet{Guriev_2019, Guriev_2020}. However, China has -- more than any other country -- perfected the strategic censorship of the internet \citep{King.2013, King.2014, Roberts_2018}, as well as modern authoritarian surveillance technologies \citep{Kostka2019, Kostka_2020, Strittmatter_2020}. Following \citet{Ringen_2016} and \citet{Minzner_2018}, we maintain that it can therefore be argued that China has taken the idea of an ``informational autocracy'' to the next level, by combining sophisticated surveillance and censorship techniques with targeted repression.}
In this new type of informational autocracy, increasingly sophisticated surveillance technologies are used to keep the population in check, but can also permit the state to conduct a new kind of industrial policy, by using data collected through surveillance as an input for research and innovation \citep{Beraja_Yang_Yuchtman_2021, Karpa_Klarl_Rochlitz_2021}. 

\begin{figure}[htbp]
\centerline{\includegraphics[scale=1]{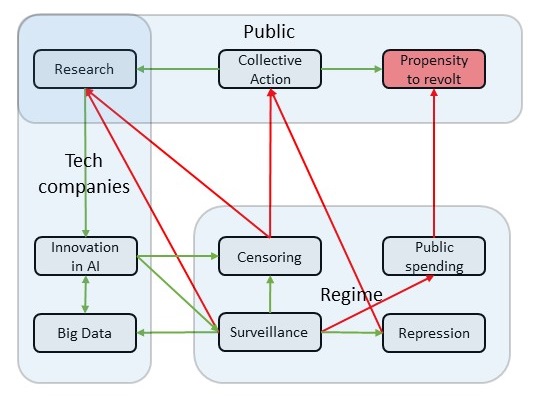}}
\caption{Theoretical Framework}
\label{fig}
\end{figure}

Figure \ref{fig} illustrates our theoretical framework, by presenting the different strategies used by authoritarian regimes in simplified form. 
Red arrows represent a negative relationship, whereas green arrows illustrate a positive effect. 
As can be seen, repression and censorship make it more difficult for the public to organize through collective action.\footnote{Whereas repression has a directly negative effect on the ability of the public to meet and protest, censorship limits the ability of the public to communicate, to access independent information, and to subsequently coordinate collective action.}
Surveillance plays a similar role, by rendering censorship \citep{Roberts.2020} as well as repression and cooptation \citep{Xu.2020} more precise and cost-effective.\footnote{The red arrow from surveillance to public spending is motivated by \citet{Pan.2020} and \citet{Xu.2020}, who show that with better surveillance, lower amounts of public spending have to be used to achieve the same reduction in the propensity to revolt.} 
As a result, all four strategies reduce the propensity of the public to revolt, and strengthen the political control of the authoritarian state.

Repression, censoring and surveillance, however, can also have direct negative effects on the ability of an economy to innovate.
While the negative effects of repression on innovation are well documented,\footnote{See for example \citet{Waldinger_2010, Waldinger_2012} and \citet{Medawar_2012} on how the expulsion of scientists from Nazi Germany in the 1930s affected the quality of research in Germany.} we focus in particular on strategies used by informational autocracies, i.e. on how censorship and surveillance might hinder research and innovation.
Section \ref{negative_effects} will discuss these negative effects in detail.

In section \ref{positive_effects}, we then turn to the effect of big data, and the potential positive interaction between authoritarian surveillance and innovation in AI.
The green arrow from surveillance to big data, and from there to innovation in artificial intelligence illustrates how data gathered by the authoritarian state can be used as an input to accelerate research in the field.
The green arrows from innovation to censoring and surveillance, in turn, symbolize the positive feedback effect of innovation in AI helping to improve censoring and surveillance technologies, which then can be used to generate even more and better data. 
After outlining both the negative and positive effects of authoritarian control on innovation in sections \ref{negative_effects} and \ref{positive_effects}, we discuss in section \ref{conclusion} under what circumstances the negative or positive effects might be more prominent.  

\section{Authoritarian Control and Innovation: Negative Effects}\label{negative_effects}

\subsection{Censorship}\label{censorship}

Censorship has been used as a control strategy by authoritarian regimes since ancient times.
In this section, we first outline why and how censorship is used in modern autocracies, before discussing how it affects innovation. 

As outlined in figure \ref{fig}, censorship can reduce the ability of the public to self-organize and revolt, both by making coordination more difficult, and by reducing the amount of information critical of the government.
Authoritarian governments, however, face a trade-off with respect to how much censorship to use.
In particular in large authoritarian states, some amount of uncensored information from institutions independent of the government might be necessary.
Otherwise, the central government might become completely dependent on information provided by regional bureaucrats \citep{Pan.2018} or the security services \citep{soldatov_rochlitz_2018}, who can use this dependence to advance their own agendas.\footnote{A detailed discussion of this trade-off for the case of Russia and China can be found in \citet{libman_rochlitz_2019}, chapter 4.}
As we will outline in more detail below, censorship also always imposes an economic cost, by rendering information more expensive. 

Authoritarian governments have reacted with different strategies to this dilemma, which are classified by \cite{Roberts_2018} according to the mechanism through which they affect the user, namely \textit{fear}, \textit{friction}, and \textit{flooding}.
Fear-based censorship operates similar to classic repression, with users being deterred from accessing or distributing certain types of information by threats of costly punishment.
Many authoritarian regimes have adopted laws and regulations that set clear boundaries regarding information usage, with journalists and content creators having to fear the possibility of arrest or loosing their jobs.
\cite{Pan.2020b} show that fear-based strategies can work, by illustrating how repression of well-known online activists in Saudi Arabia in the form of imprisonment or torture makes them less likely to continue their criticism of the government, after their release from custody.

If fear-based censorship is too obvious, however, it can produce adverse effects. 
When the arrests of online activists became public knowledge in Saudi Arabia, Google search requests for their names increased significantly (an effect often referred to as the \textit{Streisand effect}\footnote{The Streisand effect occurs when the censoring of information increases its value or makes it more attractive \citep{Hobbs.2018}. The name of the effect goes back to Barbra Streisand, who when trying to remove pictures of her home from the internet attracted even more attention to them.}), while anti-government sentiment and mobilization among the followers of the arrested activists also increased \citep{Pan.2020b}. 
In addition to the \textit{Streisand effect}, \cite{Hobbs.2018} also identify a \textit{gateway effect} of censorship.
When Instagram was blocked in China during the Hong Kong protests in 2014, users started to employ virtual private networks (VPNs), and through this then became familiar with additional politically sensitive content on other platforms that were blocked in China \citep{Hobbs.2018}. 

In order to avoid this kind of backlash, a more sophisticated form of censorship is to introduce some element of friction, which is often less easy to notice than outright repression.
Instead of restricting information explicitly, selected content is made time or resource intensive to access, rendering it more costly than other information \citep{Roberts_2018}. 
Examples of friction include the use of firewalls that block specific websites \citep{Qiang_2019}, or the manipulation of results from search engines \citep{Ruan.2017}.
Here as well, China has been a pioneer, and is more advanced today than most other authoritarian countries. 
As shown by \citet{King.2013, King.2014}, the Chinese state is not censoring all forms of government criticism from the Chinese internet.
Instead, targeted censorship is used to get rid of posts that have the potential to result in collective action against the government.
In this way, the Chinese state solves the authoritarian censorship dilemma described above, by permitting the internet to continue its role as a monitoring and information tool for the central government, while preventing it from playing the role of a ``liberation technology'', i.e. a tool that could be used by citizens to coordinate and organize public protests.

As illustrated by the green arrow from innovation to censorship in figure \ref{fig}, these kind of censorship technologies are fast evolving.
While the censorship described by \citet{King.2013, King.2014} was still mostly a manual effort by thousands of Communist party and police workers, deep learning algorithms have since started to assist and sometimes already replace manual censorship, making censorship cheaper, more sophisticated, and less easy to recognize and evade \citep{Dixon_Ristenpart_Shrimpton_2016, Roberts_2018, griffiths_2019, Tiffert_2019, Strittmatter_2020, tai_fu_2020, Gao_Qiu_Liu_2021}. 

Finally, flooding is a technique that in its results closely resembles the effect of introducing friction, i.e. access to certain types of information becomes more costly.
It differs however in the way that it is implemented, with the cost of accessing independent information being increased by the widespread distribution of alternative information congruent with the agenda of the state.
Flooding is particularly prevalent on social media, where government-affiliated actors systematically create information and distribute it.
\cite{King.2017} document how in the mid-2010s, the Chinese government was responsible for the fabrication of approximately 448 million social media posts per year. 
These posts were not used to argue with skeptics of the Communist party or the government, nor were they used to influence discussions of controversial issues, but rather to distract the public, change the subject, and altogether divert attention from discussions about politics or sensitive issues. 
In Russia, the government has been following a similar strategy, with troll factories flooding social media with pro-government messages, making it more difficult for users to identify information critical of the government within the vast mass of available content \citep{karpan_2019, linvill_warren_2020}.
As a result, while in the late 2000s social media was still described as a ``liberation technology'' and considered by many as constituting a danger to authoritarian regimes \citep{Diamond_2010}, modern informational autocracies have since reacted and seem now to be able to use social media as an additional tool to foster regime stability \citep{gunitsky_2015}. 

\medskip

\textbf{Censorship and Innovation} While fear, friction and flooding permit the authoritarian state to reduce the risk of collective action and public unrest, they also come at a cost. 
In this paper, we will focus in particular on how these censorship technologies influence the ability of an economy to generate new ideas and innovate.  

A classical example that already existed in traditional autocracies is the use of repression to get rid of scientists critical of the regime, or -- inversely -- the promotion of academics who refuse to criticize the government. 
In an excellent study of the Chinese university system, \citet{Perry.2020} shows that this still has a significant effect on scientific productivity in contemporary China, as the allocation of resources to individual researchers and projects is often based on political instead of scientific criteria. 
Similar problems were faced by the Soviet Union \citep{Graham_1987, Graham_1993}, and most importantly Nazi Germany, which suffered from an enormous brain drain as a result of the repressions of the 1930s \citep{Waldinger_2010, Waldinger_2012}.

Less obvious and more sophisticated censorship techniques can also have a negative effect on innovation, however.
One of the biggest problems faced by foreign companies in China is the way in which censorship and the great firewall are slowing down internet connections, especially to servers based abroad.\footnote{In a series of interviews with foreign investors carried out in Shanghai in May 2016 by one of the authors of this study, nearly all investors complained that internet censorship and slow connections are rendering their activities significantly more difficult. See also ``China Internet Restrictions Hurting Business, Western Companies Say'', Wall Street Journal, 12.02.2015, \url{https://www.wsj.com/articles/BL-CJB-25952.}}
Similar problems are faced by scientists and researchers.
While the domestic internet in China is easily accessible and fast, scientists who want to have unlimited access to the global web to participate in international scientific debates face significant technical hurdles and costs, as connections to the global net are slow and often many times more expensive than connections to the domestic internet.\footnote{According to an article published in The Atlantic, in 2013 Peking University charged \$1.50 a month for unlimited domestic Internet use, but \$14.50 for unlimited access to the World Wide Web (``How Internet Censorship Is Curbing Innovation in China'', The Atlantic, 22.04.2013, \url{https://www.theatlantic.com/china/archive/2013/04/how-internet-censorship-is-curbing-innovation-in-china/275188/}).}   

In addition to slow internet connections, problems with the accessibility of data also make research more difficult for scientists in China and other authoritarian countries.
One problem is that data provided by the government migth be strategically manipulated, if these data are also used as input for the performance evaluation of government officials \citep{Kalgin_2016, wallace_2016}.
\citet{Xiang_2019} maintains that in particular with respect to research in the field of artificial intelligence, data quality for China-based researchers remains heterogeneous.
While with respect to surveillance and control technologies Chinese researchers enjoy a definite advantage as compared to their peers in the West, as the security services have strong incentives to collaborate with IT firms with respect to the development of new surveillance infrastructure and the sharing of data \citep[page 45]{Xiang_2019}, this is not necessarily the case in other fields.
With respect to research in autonomous driving, speech recognition or robotics, for example, the quality and amount of available data are mixed at best, and often lower than in the US or Europe, as local and regional governments have no incentives to make data available, but on the contrary fear that providing data as an input for scientific research can get them into trouble \citep[page 32]{Xiang_2019}. 

A further issue is that especially in the social sciences, Western publishers are facing increasing pressure by the authorities in Beijing to render articles critical of the Chinese government invisible in their academic journals.
The articles are then not only not available for download, but Chinese researchers will never learn that they actually exist, if they use only the domestic version of the Chinese internet to search for information \citep{Wong_Kwong_2019}.
Similar problems exist with respect to the access to archives, to which researchers in China are only admitted if the historical information fits current political narratives \citep{Roberts_2018}.
These developments are not limited to China, with for example the Russian State Duma continuously issuing new laws that make collaboration and the exchange of information between Russian and foreign researchers more difficult,\footnote{``Russian academics decry law change that threatens scientific outreach'', Nature, 12.02.2021, \url{https://www.nature.com/articles/d41586-021-00385-5.}} a problem also faced by scientists in Hungary \citep{Enyedi_2018} or Turkey \citep{Aydin_2021}.

A common characteristic of all these censorship technologies is that they impose \textit{taxes on information}, by making accessing or sharing information more costly.
Although the primary aim of censorship is to limit anti-government activities and collective action, one externality is thus that research is becoming more costly. As shown in a seminal paper by \citet{Murray.2016}, who study a natural experiment in genetics research, a reduction in openness and available information can result in less innovation and a reduced diversity of research output.
We argue that these results can also be translated to the effect of censorship in authoritarian states, which -- through imposing taxes on information -- often has a significant negative effect on scientific research and innovation.
In a study of China's higher education sector, \citet{Schulte.2019} makes precisely this point, by arguing that novel combinations -- or innovations -- need a diverse and open academic environment to emerge.
Such an environment does only exist to a limited extent in China, where the attempt by the state to impose authoritarian stability may be in conflict with the objective of becoming an innovative knowledge economy \citep{Cao.2009, Schulte.2019}.

\subsection{Surveillance}\label{surveillance}

While imposing taxes on information is one important channel through which authoritarian institutions can negatively affect innovation, another channel is surveillance. 
Here the effect is however less straightforward.
On the one hand, surveillance can increase performance, when monitoring is inducing researchers to exert more effort. 
As we discuss in more detail in section \ref{positive_effects}, surveillance can also generate data, which -- under specific circumstances -- can be used as input to accelerate research, especially in the field of deep learning. 

However, surveillance can also have a negative effect on creativity.
This trade-off is important for the AI race between China and the US (see section \ref{intro}), with the outcome of the race depending on the still open question which type of research will ultimately be more important in artificial intelligence.
While some observers argue that big breakthroughs like the deep learning revolution are rare, and that for the foreseeable future it is the amount of available data and the resources invested in research that will determine who will win the race \citep{Spitz_2017, Sun.2017, Lee.2018}, it is also conceivable that another breakthrough will change once more the fundamentals in the field, leading to a new paradigm. 
In section \ref{positive_effects}, we argue that China has an advantage with respect to the amount of data and the ability to induce effort for average-quality researchers.
In this section, we argue that with respect to creativity, however, China's authoritarian institutions impose a tax that is similar in its negative impact to the tax on information described in section \ref{censorship}.

To illustrate what we mean by creativity, think of a maze that is characterized by having one entrance but many exits, and thus many different potential paths to take \citep{Amabile.1996}. 
When trying to find an exit, some solutions can be more elegant, or more \textit{creative}, than others.
In the face of extrinsic reward structures or surveillance, the narrowest and quickest path will be chosen to get out of the maze, and ``all behavior is narrowly directed toward attaining that goal'' \cite[page 262]{Hennessey.2003}.
In order to find a creative solution, however, immersion with the maze itself, experimentation with alternative paths, and focusing attention towards seemingly incidental aspects become important \citep{Hennessey.2003}.
Without intrinsic motivation to do the task at hand, and with an external target goal in mind, individuals will most likely rush towards the end of the maze.
As \citet{Reiss.1975} have shown, evaluation competes with task enjoyment, and attention diffuses away from the task to the fact of being evaluated. 
Or as \citet[page 262]{Hennessey.2003} puts it: ``when individuals are distracted by their excitement about a soon-to-be delivered prize or their anxiety surrounding an impending evaluation, their intrinsic motivation and enjoyment of the task at hand are undermined and they rush through their work as quickly as possible.''
Surveillance can thus lead to faster results, but at the price of less creative solutions, and a lower probability that a new fundamental breakthrough is discovered.

\medskip

\textbf{Ecosystems of Surveillance} The existence of so-called ``ecosystems of surveillance'' is one aspect in which authoritarian and democratic political systems do fundamentally differ.
Although surveillance does also exist in democracies,\footnote{In Western democracies, digital surveillance technologies are being used for example to deliver targeted campaign adds during election campaigns in what \citet{Tufekci.2014} calls ``computational politics'', or when personal data is used for commercial purposes in what has become known as ``surveillance capitalism'' \citep{Zuboff.2019}.} researchers in authoritarian systems face, on average, more surveillance both in their daily lives, and in their activity as researchers \citep{Schulte.2019, jiang_2020, Strittmatter_2020, Perry.2020, Xu.2020}.
The importance of surveillance in authoritarian states can be explained by the informational dilemma of the authoritarian government.
As a result of censorship, media control and the absence or manipulation of elections, the regime does not know the true sentiments of its citizens \citep{Edmond2013, Egorov.2020, Xu.2020}.
As a consequence, the efficient allocation of resources to co-opt regime opponents remains impossible, as the regime is uncertain which actors require co-optation, and which actors might be better controlled through repression.
Such targeted co-optation or repression is however necessary, as large-scale mass repression is only rarely used in contemporary dictatorships \citep{Guriev_2019,Xu.2020}, partially because of the disadvantages of international backlash in a globalized economy, but also because visible repression can signal that the regime is weak \citep{Guriev_2020}.

New computational methods and the ability to infer new information from existing data with the application of AI technologies have now led to a dramatic increase in digital surveillance capabilities \citep{Edmond2013, Tufekci.2014, Feldstein_2019}, which are increasingly being used by authoritarian regimes. 
One such technology are social credit systems.
While the best known system is currently being introduced in China, the technology is also exported by China to other authoritarian states \citep{Feldstein_2019}.  
The stated objective of China's social credit systems is to assess the ``trustworthiness'' of individuals by taking into account personal, financial, and behavioral data, in order to foster social control and pre-empt social instabilities \citep{Kostka_2020}.

In combination with the largest video surveillance network in the world, and the collection of biometric information on individuals such as DNA samples and fingerprints, the Chinese social credit system is in the process of becoming a comprehensive database on hundreds of millions of individuals \citep{Liang.2018, Qiang_2019}.
The system is unprecedented in its scale and scope, with the scale at which data are collected and used being made possible by the lack of a legal system protecting personal data \citep{Chen.2017}.
Since March 2020, the development of the system has received an additional boost through the compulsory introduction of tracing apps during the Covid 19 pandemic \citep{Knight_Creemers_2021}.\footnote{See also ``China uses cover of Covid to expand Big Brother surveillance and coercion'', The Times, 25.04.2021, \url{https://www.thetimes.co.uk/article/china-uses-cover-of-covid-to-expand-big-brother-surveillance-and-coercion-ndpz3klmw}.}
At least until today, the system also seems to benefit from high levels of social approval, with various studies finding that Chinese citizens are indifferent to, or even in favor of this kind of digital surveillance \citep{Kostka2019, Kostka_2020, Su2021, xu2021}.

As we will show in section \ref{positive_effects}, this extensive reservoir of personalized data can play a substantial positive role for the further development of AI technologies, should the Chinese government decide to make it systematically available to AI firms in the country. 
The Chinese social credit system is however also putting into place a massive surveillance infrastructure, providing the Chinese state with the technical ability to tightly control every aspect of the life of its citizens \citep{Strittmatter_2020, Werbach_2022}.
It is in the interest of the Chinese government to make this surveillance infrastructure as visible and present as possible in everyday life. 
While censorship becomes more effective the less visible it is (see section \ref{censorship}), surveillance relies on its ubiquitous nature to be effective \citep{Feldstein_2019}.
Behavioral modifications, or ``chilling effects'' are created even in the absence of physical violence, due to the omnipresence of surveillance and the potential subsequent assessment of behavior \citep{Feldstein_2019,Kostka2019,Kostka_2020}. 
In line with \cite{Foucault.1995}, where the panopticon disciplines inmates of a prison through a state of constant potential observation, social credit systems can induce self-monitoring and adjustment of behavior in the interest of the government  \citep{Kostka2019, Aho.2020}.

As of yet, the social and societal implications of such a large-scale surveillance project remain, however, unclear. 
In the next section, we will argue that the omnipresence of surveillance can negatively affect creativity, and through this, high-quality innovation. 

\medskip

\textbf{Surveillance, Creativity and Innovation} Creativity, that is the generation of novel and useful outcomes and ideas \citep{Amabile.2016}, is an important input for research and innovation \citep{Anderson.2014, Litchfield.2015, Acar.2019}. 
At the macro level, it is sometimes described as a basic economic input shaping technological change \citep{Mokyr.1992, Attanasi_2020}.
Some scholars have even argued that creativity is more than just an input to innovation, with the boundary between both concepts remaining fluid \citep{Anderson.2014}.
Generally, however, creativity can be understood as the generation of new ideas, while innovation is their implementation \citep{Acar.2019}.

Creative performance can suffer from a number of external constraints, such as tight deadlines and time pressure \citep{Acar.2019,Hennessey.2010}, evaluation and surveillance \citep{Amabile.1990}, and other, related factors \citep{Acar.2019}.
According to the \textit{componential theory of creativity},\footnote{For an overview of other theoretical perspectives on creativity and innovation, see \cite{Anderson.2014}.} in which domain-relevant skills, creativity-relevant skills, and intrinsic task-motivation are defined as the principal components of creative performance \citep{Amabile.1983,Amabile.1988,Amabile.2016}, the main \textit{external} influence on creativity is channeled through intrinsic task-motivation.\footnote{A related focus is proposed by \cite{Cerasoli.2014}, who review 40 years of research on external incentives and intrinsic motivation. While we focus on external \textit{constraints} and intrinsic motivation, \citet[page 983]{Cerasoli.2014} are in line with our argumentation when they show that intrinsic motivation is a strong predictor of performance, but suffers from ``crowding out'' when external incentives are too directly tied to performance outcomes.}
%We hypothesize that surveillance can undermine task motivation, and hence creativity and innovation. 
%By doing so we distinguish ourselves from others who argue that the effect of surveillance on performance is ambiguous \citep{Schedlinsky_2020}. 
%Performance, however, is distinct from \textit{creative} performance. 
%For some tasks \textit{social facilitation} may occur in the presence of others through cognitive arousal \citep{Zajonc.1965}. 
%Complex tasks, to which creative tasks form a subcategory, may not require dominant responses facilitated by cognitive arousal and thus performance should be undermined \citep{Zajonc.1965,Amabile.1990}. 
To understand how authoritarian surveillance affects creativity, one thus needs to understand how it affects intrinsic motivation.

To better understand the psychological processes of intrinsic motivation, a useful framework is provided by self-determination theory, or SDT \citep{Deci.1985,Ryan.2019}.
According to SDT, humans have three innate needs that determine motivation: the need for interpersonal relatedness, the need for competence, and the need for autonomy \citep{Ryan.2000}.\footnote{Beyond intrinsic motivation, these basic psychological needs appear to be essential for facilitating positive social development, personal well-being, and the avoidance of psychopathology \citep{Ryan.2000, Ryan.2019}.}
While interpersonal relatedness refers to being affiliated with a group or team, competence is the feeling of mastering a task, and autonomy refers to having control over a situation.
In the following, we discuss how in particular the needs of competence and autonomy, and through them intrinsic motivation, are affected by government surveillance. 

According to SDT, intrinsic motivation can either be supported or undermined by the context and external circumstances of a specific task \citep{Ryan.2000}.
In this respect, perceived levels of competence and autonomy play a particularly important role \citep{Deci.1975,Deci.1985}.
Events such as positive feedback, awards or communications which foster feelings of competence can enhance intrinsic motivation.
In a creativity-related experiment with university students, \citet{Deci.1975} for example finds that positive performance feedback led to higher levels of intrinsic motivation, whereas negative feedback decreased feelings of competence, and through this intrinsic motivation and creativity.

According to \citet[page 70]{Ryan.2000}, however, a sense of competence \textit{alone} will not enhance intrinsic motivation, ``unless accompanied by a sense of autonomy or [...] by an internal perceived locus of control''.\footnote{Other studies relying on different theoretical frameworks identify a similar importance of autonomy as catalyst to enable creative performance, see e.g. \cite{Li.2018}}
In other words, in addition to a feeling of being competent, one also has to perceive the ability to independently determine one's actions, in order for intrinsic motivation to play a role. 
In the literature, this is often paraphrased as an internal, instead of an external locus of control.\footnote{An internal locus of control means that a person is autonomously able to control her or his actions, while an external locus of control signifies that the person's actions are determined by an outside actor or institution.} 

In order to perceive the locus of control internally as opposed to externally, the \textit{functional significance} of an event is highly relevant.
According to \citet{Deci.1985}, the way a person perceives a specific context can have two functional aspects -- either a controlling one, or an informational one. 
These two aspects can be of different salience, i.e. one of them can dominate the awareness of a person, and thus be more salient. 
A context is perceived as controlling in nature if a person experiences pressure from outside to reach a specific target or goal, resulting in an external locus of control \citep{Koestner.1984}.
On the other hand, so-called ``effectance-relevant'' feedback, i.e. feedback that is perceived as providing helpful information for the achievement of a particular task, without limiting the autonomous functioning of a person, is resulting in the locus of control being perceived as internal \citep{Koestner.1984,Deci.1985}.
Here the effect on intrinsic motivation is positive. 

Experiments testing this theory have shown that imposing limits with the intention to control negatively affect intrinsic motivation and creativity, as opposed to imposing informational limits that provide feedback but are not coercive in nature \citep{Koestner.1984}\footnote{Here the difference between controlling and informational limits refers to how an external constraint of behavior is framed. When the activities of subjects were met with ``shoulds and musts'', their intrinsic motivation and creative performance were reduced. Conveying the same behavioral constraint with compassion and without external pressure had no negative effect on intrinsic motivation.}.
Similar effects were found by \citet{Pittman.1980}, who look at verbal rewards that also reveal an intention to control, or \citet{Lepper.1975}, \citet{Pittman.1980} and \citet{Plant.1985}, who study the effects of video surveillance.  
\citet{Enzle.1993} look at different forms of surveillance, and find that surveillance with the intention to evaluate,\footnote{A negative effect of evaluative surveillance has also been identified by \citet{Froming.1982}, who look at intrinsic motivation, and \citet{Amabile.1990}, who study creativity.} or to check that rules are being followed had a negative effect on motivation. 
Surveillance where these intentions were not openly stated or visible, but where the subjects suspected that either evaluation or distrust was motivating the surveillance, had an even stronger negative effect on intrinsic motivation.
If, on the other hand, subjects were convinced that surveillance was in place not to control but to assist them in their tasks, the effect on intrinsic motivation was positive.

These results are highly relevant for our study. 
As \citet[page 261]{Enzle.1993} put it, ``it is not surveillance per se that is important, but the belief that the surveillant intends to exercise social control.''
For now, it seems that even though digital surveillance is becoming increasingly omnipresent in China, it is not yet seen in a negative light by most Chinese citizens \citep{Kostka2019, Kostka_2020, xu2021} -- possibly because gamification is diverting attention from the control aspects of digital surveillance \citep{Ramadan_2018}. 
If these sentiments remain stable or will change in the future remains an open question. 
It is possible that once the intention of the Chinese state to use surveillance for social control becomes more visible, the negative effects on intrinsic motivation described above become more salient as well. 

\medskip

\textbf{Surveillance and Innovation in Chinese Academia} In Chinese research institutions and universities, this already seems to be the case. 
The Chinese education system is characterized by high levels of performance-oriented surveillance and control.
Competition to pass the national college entry exam is extremely high, and can often have a traumatizing effect on students who participate \citep{Howlett_2021}.
As students and academics are seen as a political ``risk group'', political surveillance is especially intense in Chinese universities \citep{Yan.2014, Shao_2020}, with the use of digital surveillance and social credit systems being combined with personal monitoring by senior researchers and students \citep{Perry.2020}. 
Materialistic incentives and career opportunities are offered with the specific intention to exercise political control \citep[page 14]{Perry.2020}, while video surveillance in classrooms and the ability of students to report professors in real-time via apps is used to monitor teaching activities.
Reports can lead to warnings, salary reductions or dismissal \citep[page 11]{Perry.2020}.

This system of intense surveillance is used to tightly control potential political activities of researchers and students in Chinese academia, while also being used to incentivize researchers to follow research directions indicated by the state.  
As a result, academic and scientific interests are subordinated to social stability and the interests of the Communist Party \citep{Liu.2017}, with most university professors strictly following the party line in their activities and research \citep{Hao.2016}.
This does not mean, however, that the system is not able to generate high levels of scientific output.
On the contrary, during the last 10 years the overall number of scientific publications has increased significantly in China.\footnote{``China declared world’s largest producer of scientific articles'', Nature, 18.01.2018, \url{https://www.nature.com/articles/d41586-018-00927-4}.}

Accordingly, our hypothesis -- based on the discussion above -- is that ceteris paribus, the Chinese university and research system will for some time continue to perform worse in terms of research quality than a university system in a competitive democracy, as a result of the negative effects of surveillance on intrinsic motivation and creativity.
In terms of the overall quantity of produced research, tight surveillance migth however offer an advantage, as surveillance allows to incentivize researchers to exert effort. 

For now, this hypothesis has not been tested empirically in any comprehensive way.\footnote{During the last 30 years, the Chinese bureaucratic system has proven that it can very efficiently incentivize state officials to exert effort to reach pre-determined performance criteria, such as economic growth \citep{li_zhou_2005, rochlitz_et_al_2015, jia_kudamatsu_seim_2015, yao_zhang_2015, libman_rochlitz_2019}. 
\citet{Schedlinsky_2020} however show experimentally that even for relatively simple tasks, surveillance can reduce the motivation and hence the effort exerted. 
Hence, more research is needed to better understand how surveillance can affect performance with respect to complex and less complex tasks in authoritarian environments, and if -- for example - a difference exists between performance in academic and scientific environments, and performance in bureaucratic settings.} 
It could, however, provide an explanation why China is still lagging behind with respect to research quality in the field of artificial intelligence, while it has already overtaken all other countries in the world with respect to overall research output in the field (as illustrated by Figure \ref{pubs} in section \ref{intro}). 
Table \ref{nature_index_table} provides additional evidence on this point, by showing how the top 20 research institutions in the world with respect to research quality in AI (as measured by the Nature Index of research output in applied artificial intelligence) are still mainly located in the United States.
When looking at the overall number of publications in AI, however, 14 out of the world's top 20 institutions are already located in China (Table \ref{dimensions_data_table}).

\begin{table}[]
\caption{Top 20 research institutions in artificial intelligence (Nature Index)}
\label{nature_index_table}
\resizebox{\textwidth}{!}{%
\begin{tabular}{|l|l|l|c|}
\hline
%\multicolumn{4}{|c|}{Top 20 institutions in artificial intelligence (Nature Index)}                           \\ \hline
Rank & \multicolumn{1}{c|}{Institution}                     & \multicolumn{1}{c|}{Location}  & Share 2015–2019 \\ \hline
1  & Harvard University                & United States & 331,08 \\
2  & Stanford University               & United States & 257,9  \\
3    & Massachusetts Institute of Technology   (MIT)        & United States & 209,04          \\
4  & Max Planck Society                & Germany                        & 167,98 \\
5  & University of Oxford              & United Kingdom            & 132,34 \\
6  & University of Cambridge           & United Kingdom            & 130,68 \\
7  & Chinese Academy of Sciences (CAS) & China                          & 130    \\
8  & University College London (UCL)                             & United Kingdom            & 129,7  \\
9    & Columbia University & United States & 127,56          \\
10   & National Institutes of Health (NIH)                  & United States & 122,69          \\
11 & New York University (NYU)       & United States & 117,4  \\
12 & University of Washington    & United States & 110,68 \\
13 & Princeton University              & United States & 107,92 \\
14 & University of California, Berkeley & United States & 107,54          \\
15   & University of California, San Diego & United States & 100,63          \\
16   & University of Pennsylvania                   & United States & 98,54           \\
17 & Johns Hopkins University    & United States & 85,8   \\
18 & Yale University                   & United States & 83,95  \\
19   & California Institute of Technology   (Caltech)       & United States & 83,59           \\
20 & University of Toronto & Canada                         & 83,27  \\ \hline
\multicolumn{4}{l}{Note: Data come from the Nature Index on top-level publications in the field of applied artificial intelligence, see:}\\ \multicolumn{4}{l}{\url{https://www.natureindex.com/supplements/nature-index-2020-ai/tables/countries}.}\\
\end{tabular}%
}
\end{table}

\begin{table}[]
\caption{Top 20 research institutions in artificial intelligence (Dimensions Data)}
\label{dimensions_data_table}
\resizebox{\textwidth}{!}{%
\begin{tabular}{|l|l|l|c|}
\hline
%\multicolumn{4}{|c|}{Top 100 research organizations in artificial intelligence (Dimensions data)} \\ \hline
Rank & \multicolumn{1}{c|}{Institution}                     & \multicolumn{1}{c|}{Location}  & Publications in AI\\ 
& \multicolumn{1}{c|}{}                     & \multicolumn{1}{c|}{}  & (2015 - 2019)\\ \hline
1    & Tsinghua University                           & China          & 11,867                    \\
2    & University of Chinese Academy of Sciences     & China          & 8,835                     \\
3    & Shanghai Jiao Tong University                 & China          & 8,796                     \\
4    & Beihang University                            & China          & 8,605                     \\
5    & Zhejiang University                           & China          & 8,337                     \\
6    & Harbin Institute of Technology                & China          & 8,315                     \\
7    & Nanyang Technological University              & Singapore      & 6,985                     \\
8       & University of Electronic Science and Technology of China      & China      & 6,954      \\
9    & Stanford University                           & United States  & 6,609                     \\
10   & Massachusetts Institute of Technology         & United States  & 6,314                     \\
11   & Southeast University                          & China          & 6,18                      \\
12   & University of Tokyo                           & Japan          & 6,058                     \\
13   & Huazhong University of Science and Technology & China          & 6,048                     \\
14   & University College London                     & United Kingdom & 5,988                     \\
15   & Wuhan University                              & China          & 5,797                     \\
16   & Beijing Institute of Technology               & China          & 5,641                     \\
17   & Peking University                             & China          & 5,612                     \\
18   & Harvard University                            & United States  & 5,553                     \\
19   & University of Michigan                        & United States  & 5,464                     \\
20   & Xi'an Jiaotong University                     & China          & 5,364                     \\ \hline
\multicolumn{4}{l}{Note: Data come from the Nature Index Dimensions Database that counts all publications in the field of}\\ 
\multicolumn{4}{l}{artificial intelligence between 2015 and 2019, see}\\
\multicolumn{4}{l}{\url{https://www.natureindex.com/supplements/nature-index-2020-ai/tables/dimensions-countries}.}\\
\end{tabular}%
}
\end{table}

\section{Authoritarian Surveillance and Big Data}\label{positive_effects}

Our overall hypothesis in this paper is that while the United States still have an advantage with respect to research quality and creativity, China has an advantage with respect to the number and motivation of average-quality researchers and internet entrepreneurs, as well as with respect to the quantity of data available for research.
Accordingly, who can win the AI race will ultimately depend on what will be more important in future research in AI -- big data or new ideas.
In this section, we will first give an outline of the symbiotic relationship between the Chinese state and private companies with respect to the gathering of data, before looking at how this data can be used to promote innovation. 

\bigskip

\subsection{Cooperation Between the State and the Private Sector}

While sections \ref{censorship} and \ref{surveillance} look from a political perspective at the landscape of surveillance in China, this section takes the different stakeholders into account, and shows how their cooperation permits the collection of large amounts of data on Chinese citizens. 
In China, a handful of big tech companies shape the digital environment, with Baidu, Alibaba and Tencent being the three biggest players \citep{KaiJia.2018}.
They offer a vast array of services, including online payment, e-commerce, social media, medical services, cloud services, as well as all kinds of additional convenience apps, which permit them to gather large amounts of data on their customers \citep{Arenal.2020}.
They have also developed their own social credit systems, which combine and bundle data gathered across different domains. 

From the side of the government, smart cities with surveillance technologies in the form of CCTV and a range of other sensors are being developed in metropolitan areas, with the number of surveillance cameras installed in China having increased almost exponentially in recent years.\footnote{According to a recent report by China File, central and local Chinese governments spent \$ 2.1 billion between 2016 and 2020 to buy surveillance cameras, ``State of Surveillance: Government Documents Reveal New Evidence on China’s Efforts to Monitor Its People'', China File, 30.10.2020, \url{https://www.chinafile.com/state-surveillance-china}.}
The Chinese state also places a high priority on technologies like the internet of things that are relevant for smart cities, as they facilitate economic growth but also render censorship and surveillance more effective \citep{Kshetri.2017}.

Similar trends are observable for the AI sector as a whole, where the state takes a leading role in financing AI development, buying AI solutions and determining strategies for future development \citep{Ding.2018, Arenal.2020, Righi2020}.
Through this kind of symbiosis, data are shared between public and private domains, for example in the financial sector through Alibaba's social credit system ``Zhima score''. 
Besides transaction data from more than half a billion users in Alipay, Zhima score has also access to external government data derived from interactions between individuals and public services \citep{Arenal.2020}.
The use of facial recognition software to identify deception in potential borrowers is another example from the financial sector where the state-owned Bank of China and Tencent cooperate \citep{Kshetri.2020}.
Other projects include the sharing of legal data on criminals between the authorities and private companies that provide digital service solutions such as video identification systems or remote trials on social-media \citep{Arenal.2020}.
In an attempt to automate legal decision making, courts purchase AI-powered solutions from tech companies for gathering data on court decisions \citep{Stern.2021}.
This data is subsequently made public in markets where companies and courts trade raw and repacked data, for research and commercial purposes.
Additional examples include the use of facial recognition technologies to identify jaywalkers, or smart glasses for the police, with the technology being developed by private companies that also benefit from shared data to further improve their systems \citep{Arenal.2020}. 

The sharing of large amounts of data between the state and the private sector is made possible by data privacy regulations that are much weaker than in Europe or the US \citep{Wu_Lau_Atkin_Lin_2011, Kshetri.2014}.
In particular with respect to data sharing, Chinese companies are allowed to operate with very few regulations \citep{KaiJia.2016}.
This is being facilitated by consumers being less concerned about data privacy in China than in many other countries \citep{Kshetri.2017}. 

Although the Chinese government has recently started a crack-down on the country's big tech companies, this does not mean that the symbiotic relationship between the private sector and the state is about to end. 
Rather, it can be understood as a signal by the Communist Party to the country's leading tech companies, to show who ultimately remains in charge.\footnote{``What China Expects from Businesses: Total Surrender'', The New York Times, 19.07.2021, \url{https://www.nytimes.com/2021/07/19/technology/what-china-expects-from-businesses-total-surrender.html}.} 

\subsection{Big Data and Innovation in AI}

According to \citet[page 14]{Lee.2018}, at the current stage of AI research ``successful AI algorithms need three things: big data, computing power, and the work of strong -- but not necessarily elite -- AI algorithm engineers. Bringing the power of deep learning to bear on new problems requires all three, but in this age of implementation, data is the core. That's because once computing power and engineering talent reach a certain threshold, the quantity of data becomes decisive [...].''\footnote{See also \citet{Halevy2009} and \citet{Sun.2017} on the role of big data in deep learning, especially with respect to vision tasks and facial recognition.}

\citet{Beraja_Yang_Yuchtman_2021} show that data from the symbiotic relationship between private firms and the Chinese state is already today having a measurable positive effect on the number of commercial AI innovations in China, with the positive effect being larger than the potential crowding-out effect resulting from state intervention.
The provision of government data to fuel private-sector innovation can thus be seen as an industrial policy tool used by the Chinese state, with the data gathered through surveillance being used as subsidy.

For the moment, the positive effects of this symbiotic relationship are mostly concentrated in the field of surveillance technologies \citep{Xiang_2019, Beraja_Yang_Yuchtman_2021}, where China has already today gained world leadership, and is exporting its technological solutions to other countries \citep{Feldstein_2019}.
In other sectors, China is not quite there yet, with various institutional and technological factors hindering the efficient sharing of data between government institutions and private companies \citep{Xiang_2019}.
In the health sector, for example, millions of personalized dossiers already exist, but implementation and data use are still hampered by institutional deficiencies \citep{Zhang2020}. 
Similar problems exist with respect to automated court decisions, where insufficient analytical capabilities hinder the effective implementation of the system \citep{Stern.2021}.
Even facial recognition is still suffering from implementation problems, with the efficiency of AI-automated CCTV applications remaining limited by various hard- and software problems \citep{Peterson.2020}.

Nevertheless, already today China is a leader in combining datasets from different sources and making them available for research institutions and private companies, at a much larger scale than other countries. 
Once these massive datasets are available, their use as input for deep learning algorithms can have significant transformative potential for various sectors \citep{Qiu2016}.
If one considers China's population size, it extensive and fast evolving surveillance infrastructure, as well as the fact that the country only seriously embarked on using deep learning technologies a couple of years ago, the potential for future development is vast.
If it can outweigh the  negative effects of surveillance and control on creativity and innovation remains to be seen, but is definitely a possibility. 

\section{Conclusion}\label{conclusion}

In this paper, we examine the implication of a new technology -- deep learning -- on the effect political institutions have on innovation.
Traditionally, authoritarian political institutions were seen as having a mostly detrimental effect on innovation, as they block the free flow of information, and hinder researchers to creatively explore new ideas. 
In section \ref{negative_effects}, we show that these negative effects still exist today. 
In authoritarian or semi-authoritarian countries such as China, Russia, Turkey or Hungary, the free flow of information is disrupted by various forms of censorship, and researchers face substantial pressure and control from the side of the state. 
We discuss a large theoretical and empirical literature from the field of psychology -- in particular works related to self-determination theory \citep{Deci.1975, Deci.1985, Deci.1987, Deci.1990} -- to show how the attempts of a government to exercise social control can negatively affect intrinsic motivation and creativity.
In particular in view of the recent initiative by the Chinese state to introduce a large-scale system of digitized social surveillance, these studies have again become highly relevant. 
They show that although China and other authoritarian states often have ambitious strategies to foster innovation and build knowledge economies, their authoritarian institutions can act as a significant obstacle and stumbling block in this regard. 

We then introduce a novel hypothesis, by arguing that with respect to research in artificial intelligence and in particular deep learning, the negative effects of censorship and surveillance might be attenuated -- or even outweighed -- by the positive effects of having large amounts of data available. 
As we outline in section \ref{positive_effects}, due to extensive government surveillance, lax data privacy rules and the bundling and sharing of data between state and private actors, in some sectors such as for example facial recognition, research institutions in China have already today access to much larger and better datasets than their competitors in democratic countries.
As data are the most important input in this type of research \citep{Halevy2009, Sun.2017}, the potential positive effects on future innovation might be substantial.  

For now, empirical research on these questions remains scarce. 
\citet{Beraja_Yang_Yuchtman_2021} are one of the few studies that examine in a rigorous empirical setting how the sharing of data between private and government institutions affects innovation in AI in China, with the authors finding a significant and positive effect for the sector of surveillance technologies.
To understand how deep learning technologies as well different institutional approaches with respect to data-sharing and surveillance might affect the race for leadership in artificial intelligence, we would however need much more empirical and theoretical research.
Here two lines of research are of particular importance -- how political institutions affect intrinsic motivation, creativity and innovation in contemporary academic institutions, and how the role played by big data is driving innovation in artificial intelligence. 
We hope that this study can provide a first building block for a new research agenda in this direction.

\newpage

\bibliographystyle{apalike}
\bibliography{references.bib}

\end{document}